\documentstyle[aps,twocolumn,epsfig]{revtex}

\input epsf
\newcommand{\be}{\begin{eqnarray}}
\newcommand{\ee}{\end{eqnarray}}
\begin{document}

\twocolumn[\hsize\textwidth\columnwidth\hsize\csname @twocolumnfalse\endcsname
\title{ On the Origin of the Multiplicity Fluctuations \\ in High Energy
  Heavy Ion Collisions }
 \author{G.V.Danilov and  E.V.~Shuryak}
\address{State University of New York, 
    Stony Brook, NY 11794, USA}
\maketitle
\begin{abstract}
Multiplicity fluctuations in heavy ion collisions obtain 
comparable contributions
both from initial stage of the collisions, and from final stage
interaction.
We calculate the former component, using the ``wounded nucleon'' model
and standard assumptions about nuclei and NN cross section. Combining
it with the second one, calculated previously by Stephanov,Rajagopal
and Shuryak, ref.2, we obtain good quantitative description of
experimental data (ref.3)
 from NA49 collaboration at CERN  on central PbPb collisions.
\end{abstract}
\vspace{0.1in}
]
\begin{narrowtext}   
\newpage 

\section{Introduction}
  Recently the subject of
event-by-event fluctuations have attracted
 a significant interest. On the theory side, it was motivated by
possible relation with thermodynamical observables \cite{Stod_me} 
 toward understanding, or as a background to critical fluctuations,
expected at the so called tricritical point
\cite{SRS1}. Experimentally, it was obviously stimulated by
near-perfect
Gaussian shapes of distribution  observed by NA49 experiment
at CERN \cite{NA49}.

  As emphasized in ref.\cite{SRS2},  all observables
can be divided into two broad classes: ``intensive'' (e.g. mean energy
or $p_t$ per particle, and ``extensive'' (e.g. total particle
multiplicity) ones. The latter are sensitive not only to
$final-state$ interaction effects like resonance production
(discussed in detail in \cite{SRS2}) but also to the $initial-state$
effects:
on general grounds their contributions can be comparable. The
simplest of non-statistical effects
is generated by pure geometry of the collision.
the  distribution over impact parameter b in a range
between 0 and some $b_{max}$ (depending on trigger conditions). This
particular
effect
recently discussed by Baym and Heiselberg \cite{BH}, with the
conclusion that it can account for the observed multiplicity
fluctuations.
The aim of this brief paper is to re-consider this calculation, and
also
include other non-statistical fluctuations originating at the initial
stage.  

The central quantity to be discussed is the following ratio 
\be {<\Delta N_{ch}^2> \over <N_{ch}>}\approx 2.0-2.2  \ee
where the r.h.s. is the NA49 value for 5\% centrality data to be used
in this work. If secondary particle production  be purely independent
process  govern by a Poisson distribution, the r.h.s. would be 1.
Correlations coming from resonant decays estimated in  \cite{SRS2}
lead to a value of about 1.5. The main question addressed
in this work is whether fluctuations in the  initial collision
do or do not explain the remaining part of the dispersion. 

\section{The model}

We discuss three sources of the  initial-state fluctuations:
 (i) the range of impact parameters b as already mentioned; (ii)
fluctuations in the number of participants due to {\em punched through
``spectators''}; (iii) fluctuations of the individual nucleons, or of the  
NN cross section \cite{Baym_etal}.

Nuclear distributions are parameterized as usual
\be 
n(r)={n_0 \over exp[(r-R)/a] +1}
\ee
with usual parameters for Pb, $n_0=.17 fm^{-3}$ R=6.52 fm, a=0.53 fm.
As we see below, the non-zero a is important for the (ii) component.
The probability for a nucleon to go through and become spectator is the
usual
eikonal-type $exp(-\sigma_{NN}\int_{path} dz \ n(z))$ formula. We used both the
mean $\sigma_{NN}$, or a fluctuating one\footnote{We remind the reader
  that at high energies we consider, the nucleon has no time to
  reconfigure, and so all subsequent interactions of one nucleon take
  place with the same cross section.
}. In the latter case we use
normal distribution, with the value of the
 dispersion taken from Fig.1 of \cite{Baym_etal}. For
collision energy $E\sim 100 GeV$ it is 
$\Delta \sigma_{NN}\approx .5  <\sigma_{NN}> $.
We attribute $1/\sqrt{2}$ part of this dispersion to each nucleon, as
they obviously fluctuate independently before collision.   

 We have simulated PbPb collisions and b interval
corresponding to 5\% 
of the total cross section: the resulting  distribution
of the number of participants $N_{part}$
is shown in Fig.1. Our main result (all effects included) is shown
by the
closed points:  for comparison we have also plotted 3 other
variants. If fluctuations of the cross section is switched off, we get
the distribution shown by open points. Although its shape is similar,
there is an overall shift to the maximal $N_{part}$: if nucleons do
not fluctuate, it is more difficult to punch through. Similar thing
happens if the surface
thickness
a is put to zero (the solid line). However purely geometric 
``triangular'' distribution over b, from 0 to its maximum, have a
different shape and width: it is this naive distribution which was
used in \cite{BH}. (We have found that for larger centrality cut our
results
converge toward geometric one, but for small 5\% 
cut it is clearly inadequate.) 
 
\begin{figure}[hb]
\epsfxsize=3.in
\begin{center}
\epsfig{file=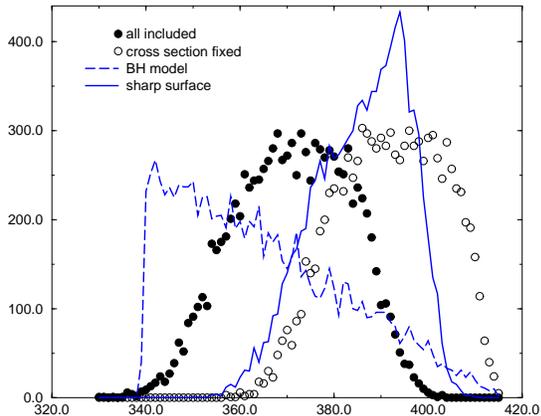,width=7.cm,angle=270}
\end{center}
\vskip -0.1in
\caption[]{
 \label{fig_1}  Distribution over the number of participants
 including all effects (closed points),  without fluctuations in
the NN cross section (open ones).  Solid line shows what happens
for  nuclei without
diffuse surface layer (a=0 in eq.(1)), while the dashed line shows the
distribution resulting from impact parameter variation, as used in \protect\cite{BH}   }
\end{figure}

 \section{Results}
Convoluting  the distribution of the number of participants 
obtained above
with the contribution of the $final-state$ interaction effects, we get the
final distribution over the observed number of charged particles $N_{ch}$:
\be
{dN \over dN_{ch}}=\int {dN_{part} P(N_{part}) \over (2\pi )^{1/2}
 \Delta N_{ch}(N_{part})}
e^{-{[N_{ch}-<N_{ch}(N_{part})>]^2 \over 2*\Delta N_{ch}(N_{part})^2}}
\ee
where the mean and dispersion are assumed to depend linearly on the
number of participants $N_{part}$. In particular, we use
\be <N_{ch}(N_{part})>= C N_{part} \ee (with C determined from the mean
observed multiplicity to be C=.75).
For the dispersion we use the estimated effect of final state
interaction
in resonance gas \cite{SRS2}, namely 
 \be \Delta N_{ch}(N_{part})^2=1.5 <N_{ch}(N_{part})> \ee

The results are shown in Fig.2. One can see that 
the distribution we obtain reproduces data rather well, 
although it is somewhat
more narrow. Let us also note, that because it is only the width
of the $N_{part}$ matters, other model distributions shown in Fig.1
do equally well. The exception is the triangular one: due to its
larger
width it is closer to data than ours. (This explains why the authors
of \cite{BH} obtained good agreement in their total width, but of
course does not justify it.)


\begin{figure}[t]
\epsfxsize=3.in
\centerline{\epsffile{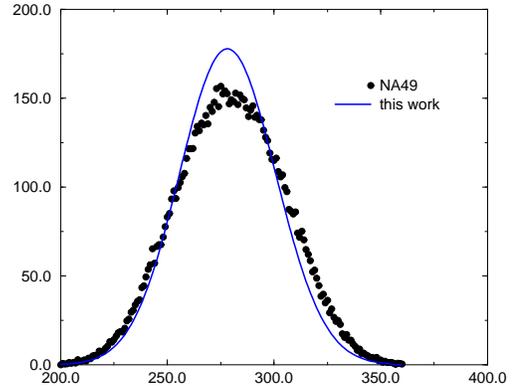}}
\vskip -0.1in
\caption[]{
 \label{fig_2} The calculated multiplicity distribution from  our
 model (solid line) is compared to the observed one
(points).  }
\end{figure}
\section{Summary, Discussion and Outlook}

In summary, we have found that initial-state fluctuations contribute about
20\% 
to the ratio (1), to be compared to 50\% from random statistics and
another 25\% 
from final state (resonance) correlations \cite{SRS2}. In sum, they
do indeed explain data at about 10\% level,
which is we think is their accuracy level. Further progress would
need much more work, including studies of the detector acceptance,
etc.

 The main physics conclusion is that
 out of three effects listed at the
beginning of section 2 the dominant one is clearly (ii), namely
the fluctuations
in a number of punched-through  spectators. In contrast to
\cite{BH}, we do not find that purely geometrical effect (i) is
important for this particular data.
 We also found fluctuations in the NN cross section (iii)
to be relatively unimportant for the width
of final distribution, adding only few percent to it.

As a discussion item, one may consider remaining
discrepancy between data and our calculation. 
Even more than the width, one may address
the origin of the asymmetry of the multiplicity
distribution, well
seen  in Fig.2: the right tail is larger than the left.
 However, before ascribing 
to this  small change of
width and/or small asymmetry any physical significance, the issue 
of acceptance should be better addressed.

For outlook, let us show that with the increasing multiplicity
expected at RHIC/LHC (both because of larger multiplicity and larger detector
coverage) the role of the initial-state fluctuations $increases$.
The ratio (1) is constructed in such a way that statistical
fluctuations always produce the same r.h.s at any multiplicity.
However, non-statistical ones we discuss do not obey it.
For example,
if
 we assume the same collision but (quite
arbitrarily) that the mean observed number of particles is 1000 (which
means C=2.7 in (4))
we get from the exactly same calculation 
\be {<\Delta N_{ch}^2> \over <N_{ch}>}\approx 2.57  \ee
to be compared with the the value about 1.9 above.

\section{Acknowledgments} 
We thank Gunter Roland for useful discussion and for multiplicity data
obtained by the NA49 collaboration we used.
This work is supported by US DOE, by the grant No. DE-FG02-88ER40388.

\end{narrowtext}

\end{document}